\RequirePackage{fix-cm}

\documentclass[smallextended]{svjour3}       

\smartqed  
\usepackage{booktabs} 
\usepackage{graphicx}
\usepackage{placeins}
\usepackage{url}
\usepackage{fancybox}
\usepackage{multirow}
\usepackage{footnote}
\usepackage{pifont}

\newcommand{\cmark}{\ding{51}}
\newcommand{\xmark}{\ding{55}}
\newcounter{observation}
\newcommand{\observation}[1]{\refstepcounter{observation}
	\begin{center}
		\Ovalbox{
			\begin{minipage}{0.93\columnwidth}
				{\bf Observation \arabic{observation}:} #1
			\end{minipage}
		}
	\end{center}
}

\makesavenoteenv{tabular}
\makesavenoteenv{table}

\makeatletter
\newcommand*{\rom}[1]{\expandafter\@slowromancap\romannumeral #1@}
\begin{document}

\title{A Deep Dive on the Impact of COVID-19 in Software Development}




\author{Paulo Anselmo da Mota Silveira Neto \and Umme Ayda Mannan \and Eduardo Santana de Almeida \and Nachiappan Nagappan \and David Lo \and Pavneet Singh Kochhar \and Cuiyun Gao \and Iftekhar Ahmed}

\date{Received: date / Accepted: date}

\maketitle

\begin{abstract}
\textbf{Context.} COVID-19 pandemic has impacted different business sectors around the world.  \textbf{Objective.} This study investigates the impact of COVID-19 on software projects and software development professionals. \textbf{Method.} We conducted a mining software repository study based on 100 GitHub projects developed in Java using ten different metrics. Next, we surveyed 279 software development professionals for better understanding the impact of COVID-19 on daily activities and wellbeing. \textbf{Results.} We identified 12 observations related to productivity, code quality, and wellbeing. \textbf{Conclusions.} Our findings highlight that the impact of COVID-19 is not binary (reduce productivity vs. increase productivity) but rather a spectrum. For many of our observations, substantial proportions of respondents have differing opinions from each other. We believe that more research is needed to uncover specific conditions that cause certain outcomes to be more prevalent.
\end{abstract}

\section{Introduction}

The latest threat to global health is the ongoing
outbreak of the respiratory disease that was recently
given the name Coronavirus Disease 2019
(COVID-19) \cite{Uncharted}. COVID-19 was recognized in December 2019, in Wuhan, a large city in central China \cite{Wuhan}. Three months later, on March 11, 2020, the World Health Organization (WHO) characterized COVID-19 as a pandemic\footnote{WHO characterizes COVID-19 as a pandemic. \url{https://www.who.int/emergencies/diseases/novel-coronavirus-2019/events-as-they-happen}}. According to WHO:\textit{``We have never before seen a pandemic sparked by a coronavirus. This is the first pandemic caused by a coronavirus. And we have never before seen a pandemic that can be controlled, at the same time.''} In August 2020, COVID-19 was spread out more than 200 countries with more than 500k confirmed deaths. 

The COVID-19 pandemic is considered the most crucial global health calamity of the century and the greatest challenge humankind faced since the 2nd World War \cite{Chakraborty}. It has enormously impacted how we live and interact with each other (social distancing, wearing masks, washing hands frequently, quarantine, etc.) Besides the health problems, the action taken by the countries around the world to manage the COVID-19 pandemic, e.g., restricting travel, shuttering nonessential businesses, and implementing universal social distancing policies, are having drastic economic consequences. In the United States, for example, more than 30 million Americans have filed initial unemployment claims since March 2020.\footnote{\url{https://edition.cnn.com/2020/04/30/economy/unemployment-benefits-coronavirus/index.html}}

Business sectors, such as travel and transportation, manufacturing, hotels, restaurants, live entertainment and movie, and sports are strongly impacted and they are having to adapt to this new situation. The software development sector is not an exception. Besides the economic aspects, suddenly, companies had to support and (in some cases) equip office workers who quickly transitioned to a work-from-home set up because of the pandemic.  

Working from home is not a new reality for software development. Many companies adopt different approaches for it, such as remote teams that have employees in a single country or even in one city, and some/all of them work from home without having to go to the office every day. Professional websites such as Upwork, Linkedin, and Stack Overflow present several job offers to work remotely. However, working from home during a pandemic is not like regular remote work. Additional difficulties are involved due to the lack of proper physical infrastructure, the need to care for children with school and daycare being closed, fear and anxiety of contracting COVID-19, etc.

Previous research studies \cite{DBLP:journals/tse/HerbslebM03}, \cite{DBLP:conf/fose/Herbsleb07}, \cite{DBLP:conf/icse/CataldoH08}, \cite{DBLP:journals/cacm/Herbsleb09} have investigated remote work and responses for a better understanding of pandemic \cite{twitter}, \cite{dashboard}, \cite{Verity}, \cite{Walkereabc0035}, \cite{covidex}. However, just a few studies \cite{bao2020does}, \cite{rahman2020exploratory}, \cite{ralph2020pandemic} have started to investigate the impact of a pandemic in software development. This aspect can be justified since COVID-19 is the first pandemic after World Wide Web development. Thus, to gain insights into the impact of COVID-19 in software development, our first research question (\textit{RQ1: What is the Impact of COVID-19 on Projects?}) explores how the pandemic impacts open source projects considering different perspectives. In addition to quantitative information about the projects, it is also important to understand the impact on developers' wellbeing; this motivates us to investigate another research question (\textit{RQ2: What is the Impact of COVID-19 on Developers WellBeing?}).

To answer these questions, we conducted a mining software repository study based on 100 GitHub projects developed in Java. The Java projects were selected according to different criteria ranging from the last update to the number of commits. We collected ten metrics for analyzing the projects. Next, we surveyed 279 software development professionals from 32 countries. The survey asked respondents to provide feedback on the impact of COVID-19 on software projects and their wellbeing.

Overall, the paper makes the following contributions:
\begin{itemize}
\item{} We perform a large scale quantitative study to investigate the impact of COVID-19 on software development based on ten different metrics. 

\item{} We complement this study with a survey of how software development professionals perceive the impact of COVID-19 on daily activities. 

\item{} Based on a set of observations from the mining software repository study and survey, we provide some recommendations for practitioners, organizations, and researchers.

\item For replication and reproducible research, we make our materials available on our project website. These include our repository mining data (project data, including metrics and time series data) and survey instrument. Our artifacts can be found at the accompanying website.\footnote{https://github.com/pamsn/covid-study}

\end{itemize}

\section{Background and Related Work}
\label{litreview}

In this section, we discuss the main work related to our study.  

\subsection{COVID-19 Studies}

The COVID-19 pandemic originated the development of several multidisciplinary initiatives around the world. The Center for Systems Science and Engineering (CSSE) at Johns Hopkins University, created an interactive web-based dashboard to visualize and track COVID-19 reported cases in real-time \cite{dashboard}. The dashboard illustrates the location and number of confirmed COVID-19 cases, deaths, and recoveries for all affected countries. Zhang et al. \cite{covidex} created, Neural Covidex, a search engine for clinicians, researchers, and other experts trying to better understand COVID-19. The system offers access to the Allen Institute for AI’s COVID-19 Open Research Dataset (CORD-19). CORD-19 is a curated public resource of more than 40,000 scholarly articles, medical reports, journal articles, and preprints about COVID-19 and the coronavirus family of viruses. Researchers from MRC Centre for Global Infectious Disease Analysis, from Imperial College London, have developed a set of tools and prediction models based on different scenarios (social distancing, shielding the elderly, and healthcare availability) \cite{Verity}, \cite{Walkereabc0035}. Chen et al. \cite{twitter} created a public Coronavirus Twitter dataset with more than 100 million tweets. According to the authors, the dataset can help track scientific coronavirus misinformation and unverified rumors, and contribute towards enabling informed solutions and prescribing targeted policy interventions.

In the software engineering area, Ralph et al. \cite{ralph2020pandemic} conducted a survey with 2225 software developers to understand the effects of the COVID-19 pandemic on developers' wellbeing and productivity. They identified that developers had lower wellbeing and productivity while working from home due to COVID-19. In addition, disaster preparedness, fear related to the pandemic and home office
ergonomics all affect wellbeing and productivity; and women, parents, and people with disabilities may be disproportionately affected. 

Rahman and Farhana \cite{rahman2020exploratory} conducted an empirical study with 129 open source COVID-19 projects hosted on GitHub to identify what categories of bugs exist in this kind of project. Initially, they identified seven categories of COVID-19 projects (aggregation, education, medical equipment, mining, user tracking, statistical modeling, and volunteer management). Next, applying open coding on 550 bug reports, they identified eight bug categories for these projects (algorithm, data, dependency, documentation, performance, security, syntax, and user interface). Based on this taxonomy, user interface was the most frequent category using the proportion of bugs across all projects, and documentation was the least frequent category.

Bao et al. \cite{bao2020does} collected data from a Chinese company, which contains the development activities from 139 developers working in 8 projects over a period of 138 working days, and compared developer productivity when working from home and working onsite. Part of records in the dataset included a period where developers worked from home due to COVID-19 pandemic. They found that working from home has both positive and negative impacts on developer productivity in terms of different metrics, such as the number of builds/commits/code reviews. In addition, they identified that working from home has different impacts on different kinds of projects. For example, working from home has a negative impact on developer productivity for large projects. They also found that the majority of developers' productivity when working from home is similar to that when working onsite. For a small portion of developers, working from home had different impacts on their productivity.

\subsection{Productivity}

There are a number of studies \cite{4273079}, \cite{8658138}, \cite{47853}, \cite{8851296}, \cite{pull},that investigate developer productivity. For example, Meyer et al.~\cite{MeyerM0F17} investigated how developer workday looks like and the relationships between their activities and perceived productivity. The study uncovered that productivity is a personal matter, and factors such as emails and meetings are often considered detrimental to productivity. In another work, Murphy-Hill et al.~\cite{Murphy-Hill19} surveyed 622 developers across 3 companies regarding self-perceived productivity and its contributing factors. They uncovered that developers perceived productivity is strongly correlated to job enthusiasm, peer support, and reception of useful feedback. Moreover, they reported that compared to other knowledge workers, {\em ability to work remotely} is more strongly related to perceived productivity. Unlike the aforementioned studies, in this work, we investigate the impact of COVID-19 on software projects.

\subsection{Wellbeing}

Many prior works investigate wellbeing in the workplace, e.g.,~\cite{BhutkarRCBAMLCG19}. However, not many investigate it in the context of software development. One such work is by Kuutila et al.~\cite{KuutilaMCEA18} that investigated the relationship between developer wellbeing and software repository metrics. They reported that developers who reported ``high hurry'' were less productive. Moreover, factors that hamper wellbeing (such as stress, sleeping problems, etc.) are negatively related to the number of chat messages. Another work by Graziotin et al.~\cite{GraziotinFWA18} investigated what happens when developers are happy and unhappy. They found that unhappiness impacts the developer's own being (in terms of low cognitive performance, mental unease, and disorder, etc.) and their work products (in terms of low productivity, low code quality, etc.). Happiness leads to the opposite effect (e.g., high cognitive performance, high productivity, high code quality, etc.).

Johnson et al. \cite{8658138} conducted a mixed-method study (two surveys and interviews) with 1159 participants from Microsoft to understand the effect of work environments on productivity and satisfaction of software engineers. They found that personalization, social norms and signals, room composition and atmosphere, work-related environment affordances, work area, and furniture, and productivity strategies were considered important factors for work environments. In addition, the ability to work privately with no interruptions and the ability to communicate with the team and leads were important factors related to satisfaction. 

Meyer et al. \cite{8666786} investigated what is a good and typical workday for software developers at Microsoft. They conducted a survey with 5971 developers and identified that on good workdays, developers make progress and value projects they consider meaningful and spend their time efficiently, with little randomization, administrative work, and infrastructure issues.  

Fucci et al. \cite{8357494} performed a quasi-experiment with 45 undergraduate students to investigate whether, and to what extent, sleep deprivation impacts the performance of novice software developers using the agile practice of test-first development (TFD). The students were divided into two groups, where 23 stayed awake the night before carrying out the tasks, while 22 slept normally. They identified that a single night of sleep deprivation reduced 50\% in the quality of the implementations. In addition, the study found that sleep-deprived developers make more fixes to syntactic mistakes in the source code.

\section{Research Design}
\label{sec:method}

\begin{center}
\begin{figure}[ht]
  \includegraphics[width=\linewidth]{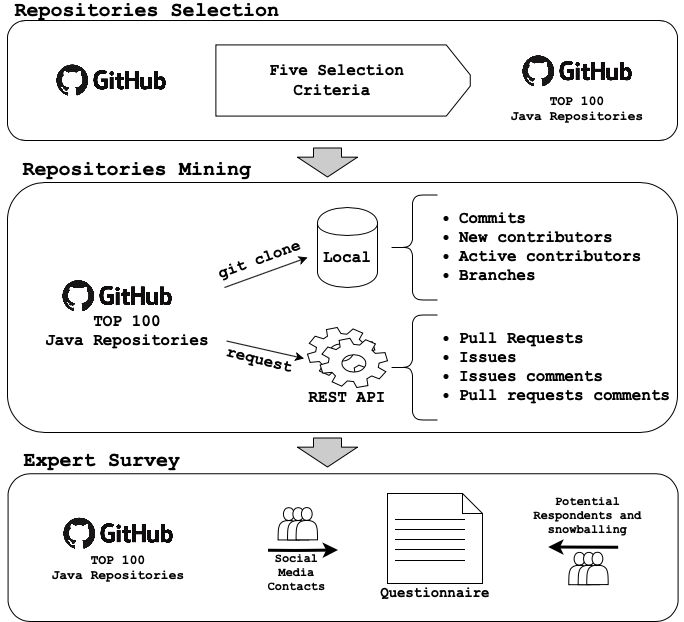}
  \caption{Research Design.}
  \label{fig:research_design}
\end{figure}
\end{center}

This section describes the research design used in our study. Firstly, a \textit{(i) Repository Mining} analysis was performed to understand the impact of COVID-19 pandemic on software projects. It considered different perspectives such as number of commits, issues, pull requests, branches, comments, time to fix a pull request and an issue, and the number of active and new contributors. Next, a \textit{(ii) Survey} with software professionals was performed to get insights on how the pandemic impact on their wellbeing. Figure \ref{fig:research_design} shows the overall research design used in our study.

\subsection{Repositories Selection and Mining}
\label{subsec:mining}

\begin{table}
\centering
\resizebox{\textwidth}{!}{\begin{tabular}{ p{4cm} p{6cm} c } 

 \hline
 \textbf{Metric} & \textbf{Description} & \textbf{Ref.} \\ 
 \hline
N of Active Contributors & The developers that had at least one commit in the study date range & \cite{4273079} \\ 
N of New Developers & The developers that committed at first time during the study date range & \cite{4273079} \\ 
N of Branches & List of all public branches of each repository & \cite{Barr2012CohesiveAI} \\ 
N of Created Pull Requests & Number of pull requests created per month considering the study date range. & \cite{Murphy-Hill19,pull} \\
N of Closed Pull Requests & Number of pull requests updated per month considering the study date range. & \cite{Murphy-Hill19,pull} \\
Pull Request created and closed per month & Number of pull requests created and closed in the same month considering the study date range.  & \cite{Murphy-Hill19,pull} \\
N of Commits & Number of commits considering the study date range & \cite{4273079} \\
N of Created Pull Requests Comments & Number of comments created related to a previous created pull request.  & \cite{7654253} \\
N of Updated Pull Requests Comments & Number of comments updated related to a previous created pull request. & \cite{7654253} \\
N of Bug-fix commits & Number of project commits related to bug fix. & \cite{123455} \\

 \hline

\end{tabular}}
\caption{Metrics used in the analysis. 
\label{tab:metrics}}
\end{table}

\textbf{Metrics.} The metrics used in our analysis are presented in Table \ref{tab:metrics}. For each of the metrics, we collected information for one year and four months (2019-Jan to 2020-May). It was important to show the behavior before and during the pandemic period. We also want to investigate how the metrics evolve month by month.

\textbf{Selecting and Filtering Repositories.} First, we choose GitHub repositories written in the Java language, since it is one of the most popular programming languages.\footnote{TIOBE Index. https://www.tiobe.com/tiobe-index/} Next, in order to detect pandemic impact on GitHub repositories, we retrieved Java repositories created from 2019 and sorted by their popularity \cite{Borges_2018}. In addition, as recommended by Munaiah et al.~\cite{Munaiah_2016}, we filtered out the noise in such large repositories by applying different inclusion and exclusion criteria, as follows: 
\begin{itemize}
    \item \textbf{Inclusion Criteria 1:} The repository has been updated at least once in the last year (2019-Jan to 2020-May);
    \item \textbf{Inclusion Criteria 2:} The repository must have at least 34 commits in the study period (2019-Jan to 2020-May); this corresponds to two commits per month in the 17-month study period. This criteria was used to filter out inactive repositories;
    \item \textbf{Inclusion Criteria 3:} The repository must have at least 10 contributors in the study period (2019-Jan to 2020-May). This criteria was used in order to eliminate irrelevant repositories, c.f., \cite{Agrawal_2018}, \cite{Krishna_2018}, \cite{Rahman_2018};
    \item \textbf{Exclusion Criteria 1:} Repositories that did not have their artifacts and description in English were not considered in the study;
    \item \textbf{Exclusion Criteria 2:} Repositories corresponding to tutorials, books, and classroom materials were also removed from our analysis.  
\end{itemize}

After the filtering step, we selected the Top 100 Java repositories. As described in Figure \ref{fig:research_design}, the repositories were cloned and the \texttt{git log} and \texttt{git branch} commands were used to get all projects commits and branches. Using the \texttt{.csv} file generated, a Python script was used to extract the number of commits, time to fix a issue and pull request, the number of new contributors (i.e. those who made their first contribution in the study period (2019-Jan to 2020-May)), the number of contributors (i.e. those who collaborated in the study period) and number of remote branches for each repository. When the information could not be collected from local cloned repositories, we used the GitHub API. Specifically, pull requests and their comments were retrieved by GitHub REST API requests\footnote{https://developer.github.com/v3/}. 

Next, we present how each metric was identified and measured for the repository mining process.

\subsubsection{Number of Active Contributors}
\textit{Active Contributors Identification and Measurement.} The number of active contributors was identified by first cloning each repository and collecting all commits\footnote{All commits were extracted using the git command: \texttt{$git log --all --format=H,aE,ci > commit_file.csv$}} in the study period. Once the commits were collected in a \texttt{.csv} file, a Python script was used to identify each developer responsible for each commit. This way, we considered an active contributor to be a developer which has at least one commit in the study period. 

\subsubsection{Number of New Developers}
\textit{New Developers Identification and Measuring.} The number of new developers was also collected from the commits extracted from each repository. The Python script searched for developers that had their first commit in the repository during the study period. It is important to know if most of the new developers initiated their contribution before or during the pandemic.  

\subsubsection{Number of Branches}
\textit{Number of Branches Identification and Measuring.} The number of branches was also collected from the cloned repositories throughout the 
GitHub REST API. It is important to mention that we used an API Endpoint that retrieves all protected and unprotected repository branches. Finally, a Python script was used to collect the branch creation date based on its commits. 

\subsubsection{Number of Created and Closed Pull Requests}
\textit{Number of Created, Closed Pull Requests Identification and Measuring.} The created and closed pull requests were collected using the GitHub REST API\footnote{\url{https://developer.github.com/v3/}} since this information is not available at local cloned repositories. According to GitHub REST API documentation, both issues and pull requests should be retrieved by the same \texttt{GET /issues} endpoint. In order to identify pull requests, we need to filter the results searching for the ``pull request'' string in each issue occurrence. With this information, we were able to identify how many pull requests were created and closed, considering the study date interval. In addition, we also identified the pull requests created and closed in the same months. With this metric, we analyze how the pandemic impacts team productivity. 

\subsubsection{Number of Commits}
\textit{Number of Commits Identification and Measurement.} The project commits were collected from the cloned repository using \texttt{git log} command. Next, the Python script was used to group all commits by month-year in a \texttt{.csv} file. This information is important to understand how the pandemic impacts on team productivity over the months. 

\subsubsection{Number of Created and Updated Pull Requests Comments}
\textit{Number of Created and Updated Pull Requests Comments Identification and Measuring.} The GitHub REST API \footnote{\url{https://developer.github.com/v3/}} was also used to collect information regarding to the pull requests comments. This metric collected both new and updated comments related to repository pull requests. The comments show important insights regarding to the project activity over the study interval. 

\subsubsection{Number of Bug-Fixing commits}
\textit{Number of Bug-Fixing commits Identification and Measurement.} To gather the number of commits related to bug-fixing, we used the GitcProc Tool\footnote{https://github.com/caseycas/gitcproc}  which uses a keyword search to determine if that commits is related to a bug fixing or not. It searches for related words such as error, bug, defect, and fix within each commit message \cite{123455}. 

\subsection{Survey}
\textbf{Protocol.} We created a 20-minute survey designed to understand the impact of COVID-19 on software development from the perspective of projects and developers' wellbeing. It was composed of seventeen closed questions on a Likert scale (Strongly Disagree, Disagree, Neutral, Agree, Strongly Agree, and N/A) and three open questions. We included the option N/A to prevent respondents providing arbitrary ratings to questions that they find to be unclear. The survey also collected demographic information from respondents. For the design of the survey, we followed the Kitchenham and Pfleeger's guidelines for personal opinion surveys \cite{BarbaraKitchenham_2008}.

We piloted our survey with three researchers (2 with Ph.D. degrees and 1 Ph.D. student) with experience in the area to get feedback on the questions and their corresponding answers; difficulties faced to answer the survey and time to finish it. As these pilot respondents were experts in the area, we also would like to know if we were asking the right questions. We conducted several iterations of the survey and rephrased some questions and removed others to make the survey easier to understand and answer. Then, we ended up with 20 questions. Another concern in this stage was also to ensure that the participants could finish it in 20-minutes. The pilot survey responses were used solely to improve the questions, and these responses were not included in the final results. We kept the survey anonymous, but in the end, the respondents could inform their email to receive a summary of the study. The survey instrument can be seen in the accompanying website\footnote{\url{https://github.com/pamsn/covid-study}}.

\textbf{Respondents}. We followed a three steps approach to recruit survey respondents: initially, we posted survey information on personal accounts of social media (Twitter, LinkedIn, Instagram). Next, most authors contacted potential respondents by email (convenience sample) and asked them to share it with other potential respondents (snowballing). Because of this process, we were not able to track the total number of invitations. In total, we received 287 responses. We disqualified six responses without actual data (without responses to any of the survey questions of interest to the study despite responding to basic demographics questions, such as the role of the respondent) and two repeated ones, leading to 279 valid responses that were considered.  

The respondents spread out in 32 countries across four continents. The top three countries where the respondents come from are Brazil, United States, and Germany. The professional experience of these 279 respondents varies from no experience to 50 years, with an average of 12.5 years and a median of 10 years.  

Regarding the size of the organizations, 34\% of the respondents work for companies with 1000 to 9999 employees, 23\% of the participants work for companies with 100 to 999 employees, 16\% of the respondents work for companies with 10 to 99 employees, 11\% of the respondents work for companies with 10000 to 99999 employees, 8\% work for companies with 100000 or more employees, and 8\% of the respondents work for companies with 0 to 9 employees. The majority of the respondents have a Bachelor's degree (41\%) and an advanced degree (41\%), i.e., Master's or Ph.D.

95\% of the respondents were paid professional, and 5\% were volunteers. 90\% of the respondents work full time, and 94\% had no disability. On average, the participants live with three people (including himself/herself), and only 32\% live with children under the age of 12. The home office experience of the 279 respondents varies from no experience to 35 years, with an average of 3.5 years and a median of 1 year. Finally, 82\% of the participants were working from office before the pandemic and switched to a home office, 17\% were working the whole time remotely, and 1\% were working at the office the whole time. 

\textbf{Data Analysis}. We collected the ratings that our respondents provided for each question. Next, we converted these ratings to Likert scores from 1 (Strongly Disagree) to 5 (Strongly Agree), We computed the average Likert score of each statement related to productivity and wellbeing perspectives and the plot Likert scale graph. A Likert scale graph is a bar chart that shows the number of responses corresponding to strongly disagree, disagree, neutral, agree, strongly agree, and N/A. In addition, we used open coding \cite{saldana2015coding}  to analyze the answers that the survey respondents gave to explain the open questions related to productivity and wellbeing, and their perceptions about how the COVID-19 pandemic has affected him/her and his/her team.

\subsection{Statistical Analysis} 

In the mining part of the study, we calculated the number of each metric mentioned in Table~\ref{tab:metrics}  for each project for $17$ months (for details, see section~\ref{subsec:mining}). To investigate the impact of COVID-19 on these metrics, we partition the time into six time-windows. As all the collected data related to project metrics are time-series \cite{book}, we did a time series analysis for each metric in each time frame across the projects to see the changes. We also performed a pairwise two-sample t-test to check if there is any statistically significant difference in the mean of each metric among the time-frames. The time series plots for each metric for each category and for each time-frame is available at the accompanying Website \footnote{\url{https://github.com/pamsn/covid-study}}.

\section{Results}

This section presents our findings from analyzing both mining data and survey responses. The content is organized around our two research questions.

\subsection{RQ1: What is the impact of COVID-19 on projects?}

Our goal is to identify the impact of COVID-19 on software projects with the first research question. For this, we conducted a mining study of $100$ Java projects where we investigate the change of $10$ different project-related metrics (defined in Table~\ref{tab:metrics}) during the COVID-19 period. In order to do this comparison, we divided our dataset of $17$ months into six time-windows:

\begin{enumerate} 
    \item \textbf{ Jan'19-May' 20 (TW1)}: This time-window includes time starting from January 2019 to May 2020.
    \item \textbf{ Jan'19-Jun'19 (TW2)}: This time-window includes the time starting from January 2019 to June 2019. 
    \item \textbf{ Jul' 19-Dec'19 (TW3)}: This time-window includes the time starting from July 2019 to December 2019. This time-window includes the initial COVID-19 spread time.
    \item \textbf{ Jan'20-Mar' 20 (TW4)}: This time-window includes time starting from January 2020 to March 2020. During this time-window, WHO (World Health Organization) declared COVID-19 as a global pandemic on March 11,2020~\cite{cucinotta2020declares}. Moreover, people started to work from home during this time-window. 
    \item \textbf{Apr' 20-May' 20 (TW5)}: This time-window includes the months of April and May 2020. 
    \item \textbf{ Jan'20-May' 20 (TW6)}: This time-window includes time staring from January 2020 to May 2020. 
\end{enumerate}

\begin{center}
\begin{figure}[ht]
  \includegraphics[width=\linewidth]{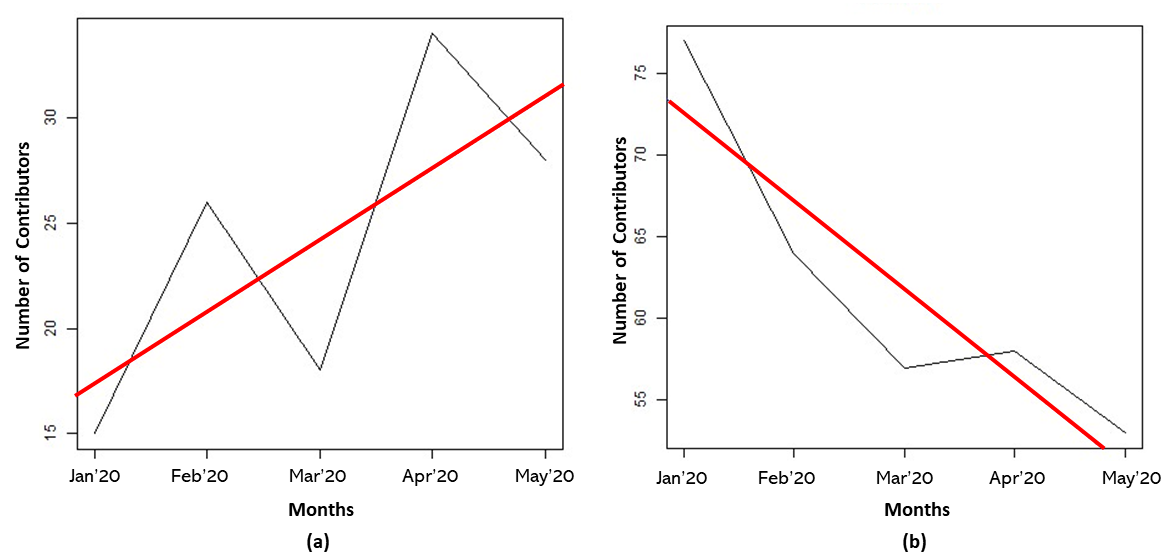}
  \caption{Month wise trends for Active Contributors across all projects.}
  \label{fig:trend}
\end{figure}
\end{center}

After dividing the metrics data into six time-windows (including the window with full-time length TW1), we investigated the changes of each metric (Table~\ref{tab:metrics}). We tracked the evolution of each metric for identifying their trends in the projects during the time-windows. For this purpose, we calculated the effect size of the month on each metric using a linear regression model, giving us how much that metric changed for each project per month. Then, based on the effect size, we categorized each project into one of two categories: increasing or decreasing. If the effect was positive, we marked those projects as increasing for that metric. For a negative effect, we marked the project as decreasing. For example, Figure~\ref{fig:trend} shows that for the metric ``number of active contributors'', project ``Brave'' (Figure~\ref{fig:trend}.a) shows an increasing trend for the time-window January 2020 to May 2020 and ``ExoPlayer'' project  (Figure~\ref{fig:trend}.b) shows a decreasing trend according to our time series analysis. All the figures generated from the analysis are provided in the accompanying website\footnote{\url{https://github.com/pamsn/covid-study}}. In Table~\ref{tab:timeseries}, we report the number of projects belonging to each trend for the six time-windows mentioned earlier.

\begin{table}[hbt]
\centering
\caption{Summary of time series analysis for each category }
\label{tab:timeseries}
\begin{tabular}{|l|c|c|c|c|c|c|c|c|c|c|c|c|}
\hline
\multirow{2}{*}{Metric}                                                                   & \multicolumn{2}{l|}{TW 1} & \multicolumn{2}{l|}{TW 2} & \multicolumn{2}{l|}{TW 3} & \multicolumn{2}{l|}{TW 4}  & \multicolumn{2}{l|}{TW 5}& \multicolumn{2}{l|}{TW 6} \\ \cline{2-13} 
& $\uparrow$  & $\downarrow$ & $\uparrow$ & $\downarrow$  & $\uparrow$  & $\downarrow$  & $\uparrow$   & $\downarrow$  & $\uparrow$  & $\downarrow$ & $\uparrow$  & $\downarrow$ \\ \hline
\begin{tabular}[l]{@{}l@{}}Active \\ Contributors\end{tabular}                               & 42               & 58              & 42               & 58              & 45               & 55                  & 31               & 69          	& 47               & 50              & 18               & 82              \\ \hline
\begin{tabular}[c]{@{}l@{}}New \\ Developers \end{tabular}														                         & 49               & 49              & 41               & 59              & 29               & 69                  & 30               & 66          	& 49               & 41              & 18               & 70              \\ \hline
Branches                                                                                     & 93               & 7               & 26               & 28              & 35               & 39                  & 91               & 9           	& 33               & 29              & 77               & 23              \\ \hline
\begin{tabular}[c]{@{}l@{}}Created \\ Pull\\ Requests\end{tabular}                             & 81               & 11              & 29               & 25              & 35               & 35                  & 73               & 15          	& 49               & 21              & 54               & 30              \\ \hline
\begin{tabular}[c]{@{}l@{}}Closed \\ Pull\\ Requests\end{tabular}                              & 45               & 54              & 45               & 53              & 49               & 50                  & 34               & 64          	& 49               & 48              & 25               & 71              \\ \hline
\begin{tabular}[c]{@{}l@{}}Pull request \\opened \& \\closed in \\same month\end{tabular} & 46               & 53              & 42               & 58              & 44               & 55                  & 39               & 58          	& 52               & 43              & 20               & 75              \\ \hline
Commits                                                                                      & 45               & 55              & 40               & 60              & 44               & 56                  & 32               & 68          	& 55               & 44              & 23               & 77              \\ \hline
\begin{tabular}[c]{@{}l@{}}Created \\Pull request \\ Comments \end{tabular}                  & 56               & 41              & 36               & 64              & 38               & 46                  & 34               & 55          	& 50               & 34              & 23               & 59              \\ \hline
\begin{tabular}[c]{@{}l@{}}Updated \\ Pull request \\ comments \end{tabular}                 & 58               & 39              & 38               & 62              & 37               & 45                  & 38               & 51          	& 49               & 35              & 32               & 51              \\ \hline
\begin{tabular}[c]{@{}l@{}} Bug-fix \\ Commits\end{tabular} 								 & 40    			& 60			  & 38      		 & 62    		   & 37				  & 62					& 32  			   & 65   			& 49               & 47				 & 25				& 70   			  \\ \hline
\end{tabular}
\end{table}

In the following sections, we list the observations pertaining to the impact of COVID-19 on different project activities.
\subsubsection{Impact on Bug-fixing Activity}

From our mining analysis, we found that in general, more projects are showing decreasing trends in the number of bug-fix commits per month during the pandemic period. For example, time-window between Jan'20-May'20 (TW6) shows more projects with a decreasing trend in comparison to the time-windows between Jan'19-Dec'19 (See Table~\ref{tab:timeseries}). Next, we performed a pairwise two-sample t-test to check if there is any statistically significant difference in the mean between these time-windows and found that between the time-windows Jul'19-Dec'19 and Jan'20-May'20, there is a statistically significant difference in their mean (p-value $<$ 0.05). Table~\ref{tab:hypotest} shows the summary of the pair-wise t-test for each metric in different time-windows. Table~\ref{tab:hypotest} only shows the statistically significant results of our t-test.

\begin{table}[hbt]
\centering
\caption{Summary of t-test among projects between different time range.}
\label{tab:hypotest}
\begin{tabular}{|l|l|l|r|r|r|}
\hline
Metric                  & TW 1 & TW 2 & Mean1 & Mean2  & p-value \\ \hline
Active Contributors     & Jul19-Dec19 & Jan20-May20 & 19.13  & 29.02 & 0.019   \\ \hline
Closed Pull Requests    & Jan20-Mar20 & Apr20-May20 & 78.79 & 46.94 & 0.047   \\ \hline
Created Pull Requests   & Jul19-Dec19 & Jan20-May20 & 6.79  & 12.48 & 0.04    \\ \hline
New Developers           & Jan20-Mar20 & Apr20-May20 & 8.88 & 4.46   & 0.00012 \\ \hline
\begin{tabular}[l]{@{}l@{}}Pull Request opened \& \\closed in the same month \end{tabular} & Jan20-Mar20 & Apr20-May20 & 62.25   & 35.7  & 0.048   \\  \hline
Bug-fix Commits           & Jul19-Dec19 & Jan20-May20 & 69.92 & 32.54   &  0.028 \\
\bottomrule
\end{tabular}
\end{table}

In our survey, we asked respondents if they agree/disagree with the statement, ``Since you began working from home, the number of bug-fixes is lower than usual''. Among the 279 survey respondents, 44.08\%  disagreed, where 15.77\% agreed that the number of bug-fixes is lower than usual since they began working from home during the pandemic. 42, 81, and 95 respondents strongly disagree, disagree, and neutral with this statement. The average Likert score for this statement is 2.58 (i.e., between "disagree" and "neutral").

\observation{Though the majority of projects show decreasing trends in the number of bug-fix commits most of the survey respondents disagree regarding decreased number of bug-fix commits during the pandemic period.}
We also asked the respondents about their view on the statement: ``The number of bugs in the project has increased since they began working from home''. 39, 91, and 107 respondents strongly disagree, disagree, and neutral with this statement. The average Likert score for this statement is 2.47 (i.e., between "disagree" and "neutral"). The following are some comments that refute or confirm the statement:

\xmark \textit{"Overall we've been finishing more features at a higher quality.}" 

\cmark \textit{"The overall level of engagement has increased, but I feel that the quality has suffered somewhat.}" 

\observation{46.60\% of the respondents conveyed that the number of bugs did not increase during the pandemic period.}

\begin{center}
\begin{figure}[ht]
  \includegraphics[width=\textwidth]{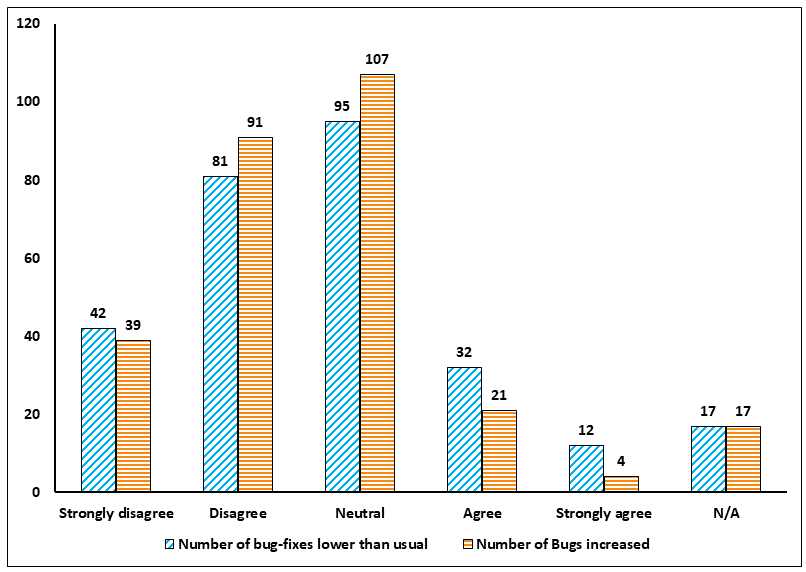}
  \caption{Summary of the survey results on the impact of COVID-19 on the bug-fixing activities.}
  \label{fig:bug-fix}
\end{figure}
\end{center}

Figure~\ref{fig:bug-fix} shows the summary of the survey results on the impact of COVID-19 on the bug-fixing activities.

\subsubsection{Impact on Discussion}
From our mining part of the study, we found that in case of creating and updating pull request comments (which is a way of discussion among developers), more projects are showing decreasing trends during the pandemic period. For example, time-windows between Jan'20-May'20 (TW6) show more projects with a decreasing trend in comparison to the time-windows between Jan'19-Dec'19 (See Table~\ref{tab:timeseries}) in both creating and updating comments. Next, we performed a pairwise two-sample t-test to check if there is any statistically significant difference in the mean between these time-windows and did not find any statistically significant difference in mean between any two time-windows (p-value $>$ 00.5, see Table~\ref{tab:hypotest} for details).

We asked our survey respondents about the review activity (another form of discussion among developers) in their projects during the pandemic. 54.83\% of the survey respondents mentioned that, in general, review activities are not lower than usual during the pandemic period. 56, 97, and 60 respondents strongly disagree, disagree, and are neutral with this statement, respectively. The average Likert score for this statement is 2.41 (i.e., between "disagree" and "neutral"). The following are some comments that refute the statement:

\xmark \textit{"I having more structure timed schedule. Review with the team, coding, review with the team.}"

We also asked about the rate of discussion among team members in our survey. 37.63\% respondents did not think there is an increment in discussion among team members since they began working from home in the pandemic period. 19, 86, and 73 respondents strongly disagree, disagree, and are neutral in this, respectively. The average Likert score for this statement is 3.0 ("neutral"). The following are some comments that refute or confirm the statement:

\xmark \textit{"Normally during work days, at least on the team/office I worked with/on we would have some time to discuss about problems that occur to us during the development of some task. Now with everyone so distant, even though we have tools to workaround that problem it still feels like you are significantly distant from that person, both physically and mentally.}"

\cmark \textit{"On our company we have a permanent virtual room opened to attend doubts and have discussions.}"

\observation{Though the number of pull request comments created and updated decreased over the pandemic period, discussion among developers related to their work and review activity did not decrease during the pandemic.}

53.04\% of the respondents disagree on the quality degradation of discussion among team members since they began working from home, while 26.16\% agree that there is a degradation in discussion quality. The average Likert score for this statement is 2.63 (i.e., between "disagree" and "neutral"). The following are some comments that refute or confirm the statement:

\xmark \textit{"I make calls with my colleagues the whole time, to discuss and cheap talk, this helps focus and fell like we are at the same room.}"

\cmark \textit{"In general, people are more nervous and stressed, which makes discussions less constructive and more stressful.}"\\

\begin{center}
\begin{figure}[ht]
  \includegraphics[width=\textwidth]{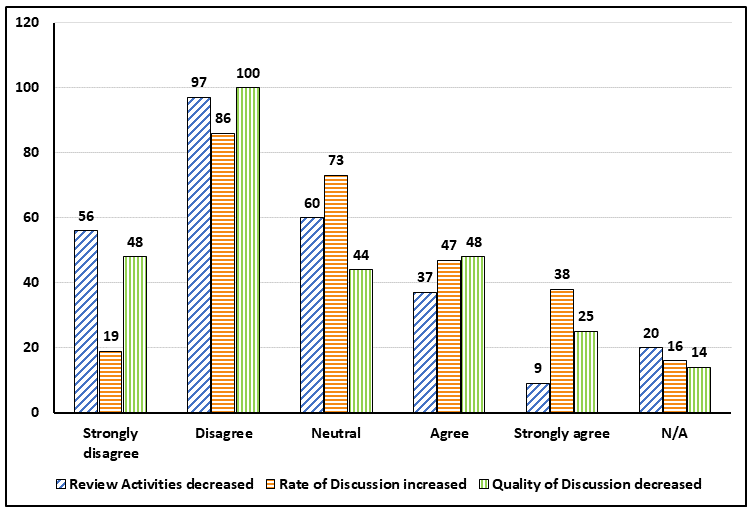}
  \caption{Summary of the survey results on the impact of COVID-19 on discussions among developers.}
  \label{fig:discussion}
\end{figure}
\end{center}

Figure~\ref{fig:discussion} shows the summary of the survey results on the impact of COVID-19 on discussion among developers.

\subsubsection{Impact on Code Contribution}
To investigate the impact of COVID-19 in code contribution, we analyzed the number of pull requests created, closed each month during and before the pandemic period. 

From our mining analysis, we found that in general, more projects show increasing trends in the number of created and closed pull requests per month during the pandemic period. For example, time-windows between Jan'20-May'20 (TW6) show more projects with an increasing trend in comparison to the time-windows between Jan'19-Dec'19 (See Table~\ref{tab:timeseries}). Next, we performed a pairwise two-sample t-test to check if there is any statistically significant difference in the mean of the number of created pull requests between these time-windows and found that between the time-windows Jul'19-Dec'19 and Jan'20-May'20, there is a statistically significant difference in their means (p-value $<$ 0.05). For the closed pull requests, we found that between the time-windows Jan'20-Mar'20 and April'20-May'20, there is a statistically significant difference in their means (p-value $<$ 0.05, see Table~\ref{tab:hypotest} for details).

We also found that the pull requests that opened and closed in the same month show increasing trends in more projects during the pandemic compared to the time-windows before the pandemic. We found a statistically significant difference in mean for this metric between the time-windows Jan'20-Mar'20 and Apr'20-May'20.

In our survey, we asked our respondents about code contribution. There are different ways of contributing code, such as adding a new feature, refactoring, etc. 58.42\% of our survey respondents do not think that working from home during pandemic results in adding fewer new features than usual time. 49, 114, and 54 respondents strongly disagree, disagree, and are neutral with the statement: ``The number of new features is lower than usual'' statement, respectively. The average Likert score for this statement is 2.41 (i.e., between "disagree" and "neutral"). The following are some comments that refute the statement:

\xmark \textit{"I am doing home office. It's effective to keep the features delivery.}"

\xmark \textit{"Overall we've been finishing more features at a higher quality. However, we have limited QA resources (we only have one full-time QA staff member), so the fact that the developers are churning out code is probably making it much more stressful for our QA lead.}"


48.75\% of our survey respondents disagree that the number of refactoring is lower than usual since they began working from home. 36, 100, and 92 respondents strongly disagree, disagree, and are neutral with this statement, respectively. The average Likert score for this statement is 2.50 (i.e., between "disagree" and "neutral"). The following are some comments that confirm the statement:

\cmark \textit{"Besides of interpersonal communication be affected, I perceived that long term strategies about the quality and refactoring, for example, are depreciated instead of short term tasks, such as new implementations and tests automation related to the implementations.}"

\observation{In general, working from home during pandemic period does not impact code contribution.}

\begin{center}
\begin{figure}[ht]
  \includegraphics[width=\textwidth]{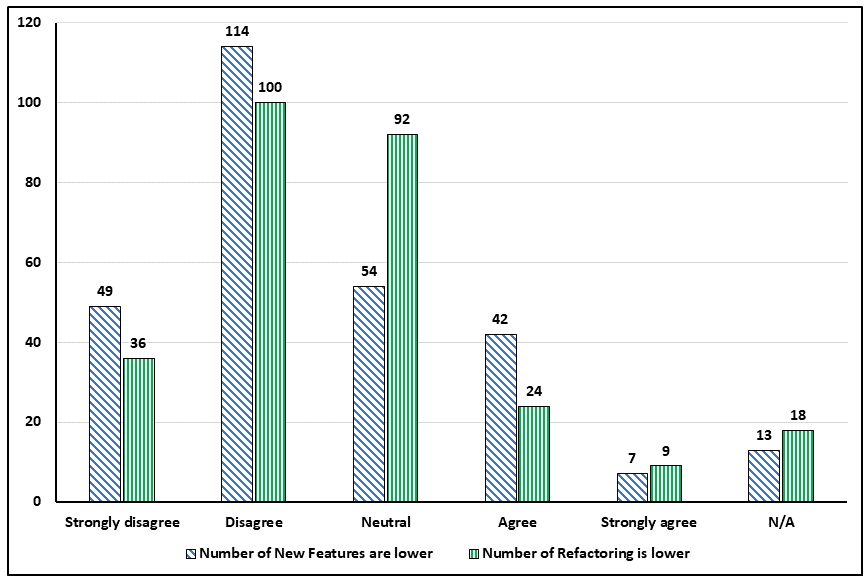}
  \caption{Summary of the survey results on the impact of COVID-19 on code contribution.}
  \label{fig:code_cont}
\end{figure}
\end{center}

Figure~\ref{fig:code_cont} shows the summary of the survey results on the impact of COVID-19 on code contribution.

In our survey, we also asked the survey respondents about task completion time, productivity, and quality of their work during the pandemic period. 

47\% survey respondents disagree that it takes longer to complete the tasks than usual since they began working from home. 58, 73, and 52 respondents strongly disagree, disagree, and are neutral with this statement, respectively. The average Likert score for this statement is 2.72 (i.e., between "disagree" and "neutral"). The following are some comments that refute or confirm the statement:

\xmark \textit{"I've more time to study my tasks, with more attention and patience, I've created step by steps before to start working and this is more effective.}"

\xmark \textit{"We scheduled more small conversations during the day and splitted more the tasks. We are keeping the same productivity as before or even higher.}"

\cmark \textit{"Tried using the Pomodoro method of time boxing tasks. Otherwise, nothing. I don’t think it helped much.}" 

\cmark \textit{"Work more hours to complete necessary work, try to be more organized with tasks and time available (prioritization). Not so effective, I still have to work more hours.}"\\


On the overall productivity, 56\% of the survey respondents disagree they are less productive since they began working at home due to the COVID-19 pandemic. 89, 67, and 39 respondents strongly disagree, disagree, and are neutral, respectively. The average Likert score for this statement is 2.43 (i.e., between ``disagree'' and ``neutral''). The following are some comments that refute or confirm the statement:

\xmark "\textit{My productivity at work increased a lot in this pandemic and I was even recognized as an QA MVP (Most Valuable Player) for the last 2 months. However, I had to keep track of my working hours because I started working a lot more than I was supposed to.}"

\xmark "\textit{My team became more productive due to the absence of office distractions.}"

\cmark "\textit{We had been working remotely for years already. However, our productivity still took a big hit - not due to working from home specifically, but due to the cognitive overload we're all experiencing these days. It's hard to concentrate and be productive when we're all worried about our families, our society and our own health.}"

\cmark "\textit{Decreased one-on-one conversations, low internet bandwidth, irregular power supplies and many other factors have impacted our effectiveness.}"\\

In general, respondents disagree or are neutral considering that the quality of their work is lower than it should have been since they began working from home. 83, 89, and 40 respondents strongly disagree, disagree, and are neutral with this statement, respectively. The average Likert score for this statement is 2.33 ("i.e., between "disagree" and "neutral").

\observation{Overall productivity and task completion time do not decrease during the pandemic period compare to the usual time.}

\begin{center}
\begin{figure}[ht]
  \includegraphics[width=\textwidth]{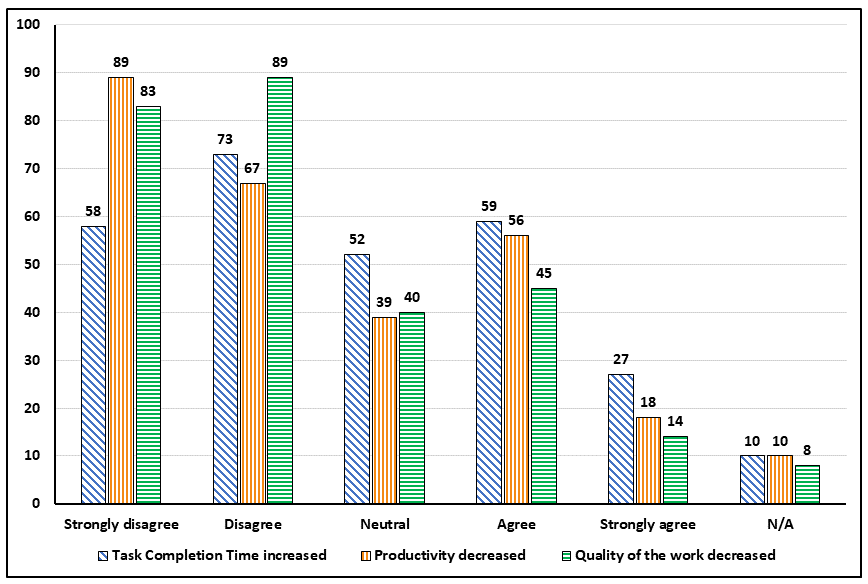}
  \caption{Summary of the survey results on the impact of COVID-19 on task completion time, productivity, and quality of the work.}
  \label{fig:task_comp}
\end{figure}
\end{center}

Figure~\ref{fig:task_comp} shows the summary of the survey results on the impact of COVID-19 on task completion time, productivity, and quality of the work.

In general, respondents disagree or are neutral, considering that the amount of testing is lower than usual since they began working from home during the pandemic. 54, 110, and 69 respondents strongly disagree, disagree, and are neutral with this statement, respectively. The average Likert score for this statement is 2.28 (i.e., between ``disagree'' and ``neutral''). The following is a comment that confirms the statement:

\cmark \textit{"Besides of interpersonal communication be affected, I perceived that long term strategies about the quality and refactoring, for example, are depreciated instead of short term tasks, such as new implementations and tests automation related to the implementations.}"\\


Many respondents disagree or are neutral, considering that the code quality has decreased since they began working from home. 62, 112, and 66 respondents strongly disagree, disagree, and are neutral with this statement, respectively. The average Likert score for this statement is 2.21 ("i.e., between "disagree" and "neutral"), which is the lowest among all survey questions. 

\observation{According to our survey respondents working from home during pandemic period does not impact code quality.}

\begin{figure}[ht]
\centering
  \includegraphics[width=\linewidth]{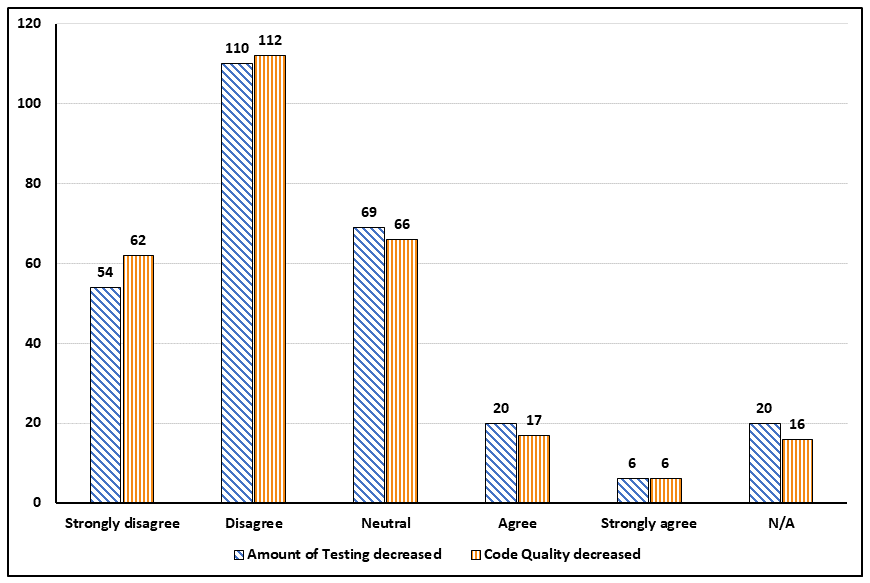}
  \caption{Summary of the survey results on the impact of COVID-19 on code quality.}
  \label{fig:code_qual}
\end{figure}

Figure~\ref{fig:code_qual} shows the summary of the survey results on the impact of COVID-19 on code quality.

\subsection{RQ2: What  is  the  impact  of  COVID-19  on  Developers  Wellbeing?}

With the second research question our goal is to identify the impact of COVID-19 on developers wellbeing. To achieve this goal, we asked the survey respondents regarding their stress, emotional condition, getting help from others, etc. In the following sections, we present our findings regarding the impact of COVID-19  on developers wellbeing.

\subsubsection{Sleep Disorder}

Sleep disorder could hamper developer's productivity. In our survey, we asked the respondents if they think their sleeping disorder increased during the pandemic period. Among 279 respondents, 39.40\% of the respondents agreed that their sleep disorder increased since they began working from home. The average Likert score for this statement is 3.0 (i.e., "neutral"). The following are some comments that refute or confirm the statement:

\xmark \textit{"This situation is quite complex, especially when we perceive a complete neglect on the part of the federal authorities in Brazil with the pandemic and at the same time see the increasingly devastating advance of the virus in Brazilian territory. The most important action has been to control the amount of information to maintain emotional control. Apparently just for sleep it hasn't been enough.}"

\cmark \textit{"I'm going to sleep at the same time. It was effectively strong.}"

\cmark \textit{"I gave sleep a priority. Discontinues some non-essential activities.}"

\observation{39.40\% of the survey respondents agreed that sleep disorder increased during pandemic period.}

\subsubsection{Level of Stress}

We found that 141 out of 279 survey respondents agreed that the level of stress increased in the pandemic period since they began working from home. A substantial number of respondents choose to be neutral or disagree (44 neutral respondents, and 85 respondents who disagree or strongly disagree). The average Likert score for this statement is 3.33 ("i.e., "neutral") which is the highest among the survey questions. The following are some comments that refute or confirm the statement:

\xmark \textit{"Constant breaks (10minutes every 50 minutes). No overtime. When my workday is done, I close my laptop and put in a drawer ("Out of sight, out of mind") and get it out only on the next working day. I feel less stressed because I have more time overall (no commuting).}"

\xmark \textit{"The pandemic is serious, but it is far away to be my source of stress. I have low need for social interaction. Sunbathing with the baby is enough to relieve the "quarantine".}"

\cmark \textit{"Working more flexible hours to watch family more during day and shift more at night. This has maintained work output and quality but increases stress.}"

\cmark \textit{"In general, people are more nervous and stressed, which makes discussions less constructive and more stressful.}"
\observation{50.53\% of survey respondents indicate that their stress level increased since they began working from home during pandemic.}

\subsubsection{Happiness}

Many respondents disagree or are neutral considering that the level of happiness increased since they began working from home. 27, 70, and 95 respondents strongly disagree, disagree, and are neutral with this statement, respectively. The average Likert score for this statement is 2.91 ("i.e., between "disagree" and "neutral"). The following are some comments that refute or confirm the statement:

\xmark \textit{"I am depressed and taking anti-depression pills along with therapy. I check a doctor online.}"

\xmark \textit{"Slightly increased the frequency of my panic attacks.}"

\xmark \textit{"I try to get just enough news to stay informed but other than that I try to avoid news because  the more I hear, the more depressed I get. I have a regular schedule of zoom dates with my boyfriend on top of our impromptu calls/facetimes to make sure that we are staying as connected as we can. That helps to some extent because we know we'll get through it, but it still completely sucks that we don't get to see each other in person. To make up for it, we probably text more frequently during work hours than we would have before the pandemic. At work, my team has weekly 30min meetings to just hangout together on zoom. They were good at first to help us stay connected and sane, but now I think we're kind of getting bored with them. We probably don't need them so frequently.}"
 
\cmark \textit{"My overall wellbeing was not affected. Just the mood and motivation is down due to the situation. There is not much to do about it. We just have to get through it.}"

\observation{34.76\% of the survey respondents mentioned that the level of happiness did not increase during the pandemic period.}

\begin{center}
\begin{figure}[ht]
  \includegraphics[width=\textwidth]{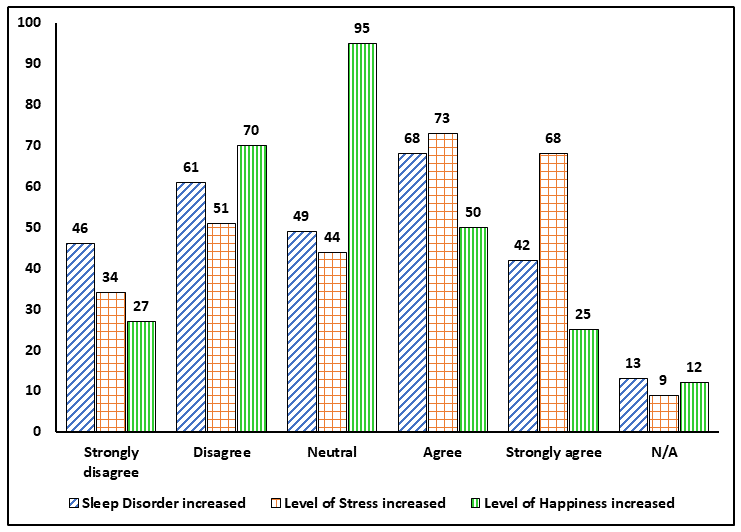}
  \caption{Summary of the survey results on the impact of COVID-19 on the developer's wellbeing.}
  \label{fig:well-being1}
\end{figure}
\end{center}

Figure~\ref{fig:well-being1} shows the summary of the survey results on the impact of COVID-19 on developer's sleep, stress level and happiness.

\subsubsection{Mentoring}

101 out of 279 survey respondents agreed that the mentoring activities for newcomers has decreased in the project since they began working from home in pandemic. A substantial number of respondents choose to be neutral or disagree (66 neutral respondents, and 81 respondents with disagree or strongly disagree with this statement). The average Likert score for this statement is 3.15 (i.e., "neutral"). The following are some comments that confirm the statement:

\cmark "\textit{I think it affected the welcoming of the youngest to the team. Face-to-face activities generate more intimacy between people.}"

\cmark "\textit{The COVID-19 has influenced the acceptance of new employees. The efficiency of the discussion among team members was much lower than face-to-face discussion.}"

\observation{35.12\% of the survey respondent mentioned that mentoring activities for newcomers have decreased during the pandemic period.}

\begin{center}
\begin{figure}[ht]
  \includegraphics[width=\textwidth]{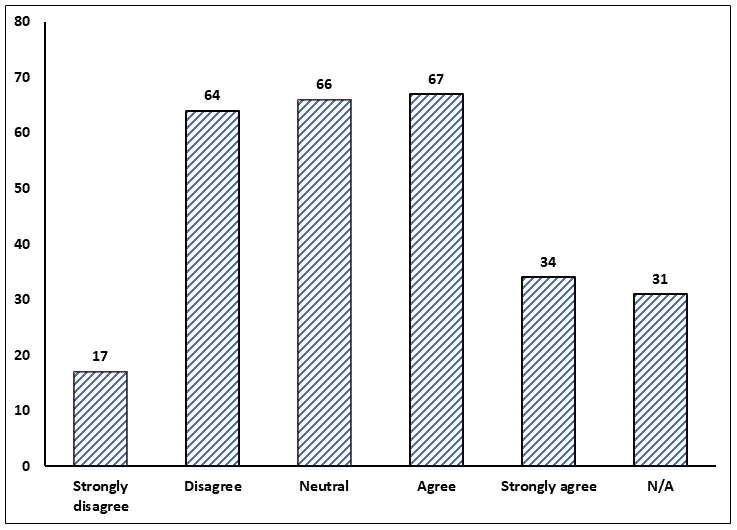}
  \caption{Summary of the survey results on the impact of COVID-19 on the developer's mentoring activity.}
  \label{fig:well-being2}
\end{figure}
\end{center}

Figure~\ref{fig:well-being2} shows the summary of the survey results on the impact of COVID-19 on the developer's mentoring activity.

\subsubsection{Interruptions}

In general, respondents disagree or are neutral considering that newly introduced interruptions along with the old ones are negatively impacting their productivity since they began working from home. 35, 81, 64 respondents strongly disagree, disagree, and are neutral with this statement, respectively. The average Likert score for this statement is 2.87 (i.e., between "disagree" and "neutral"). The following are some comments that refute or confirm the statement:

\xmark \textit{"Working remotely, others can't interrupt my work so quickly, so this increased my productivity in general.}"

\xmark \textit{"Home office leads to much less interruptions. This helps productivity a lot!}"

\cmark \textit{"I had to improvise a home office in my daughter's room so I can isolate myself there to avoid too many interruptions or background noises. It helped a little, but my wife and daughter still eventually interrupt me during my working hours.}"

\cmark \textit{"We tried to schedule a series of video conferences to keep in touch with the team. It helped a lot, but is not as good as talking directly to the persons. The rate of interruptions did increase.}"\\

\observation{41.60\% of the survey respondents do not think that newly introduced interruptions along with the old ones are negatively impacting their productivity during the pandemic time.}

We posit that interruptions are associated with number of people living in the household. We analyzed the results from the survey and our results show that the average Likert score ranges from $2.9$ to $3.4$, indicating the perception that the number of people living in the household is not having a major impact.

\begin{center}
\begin{figure}[ht]
  \includegraphics[width=\linewidth]{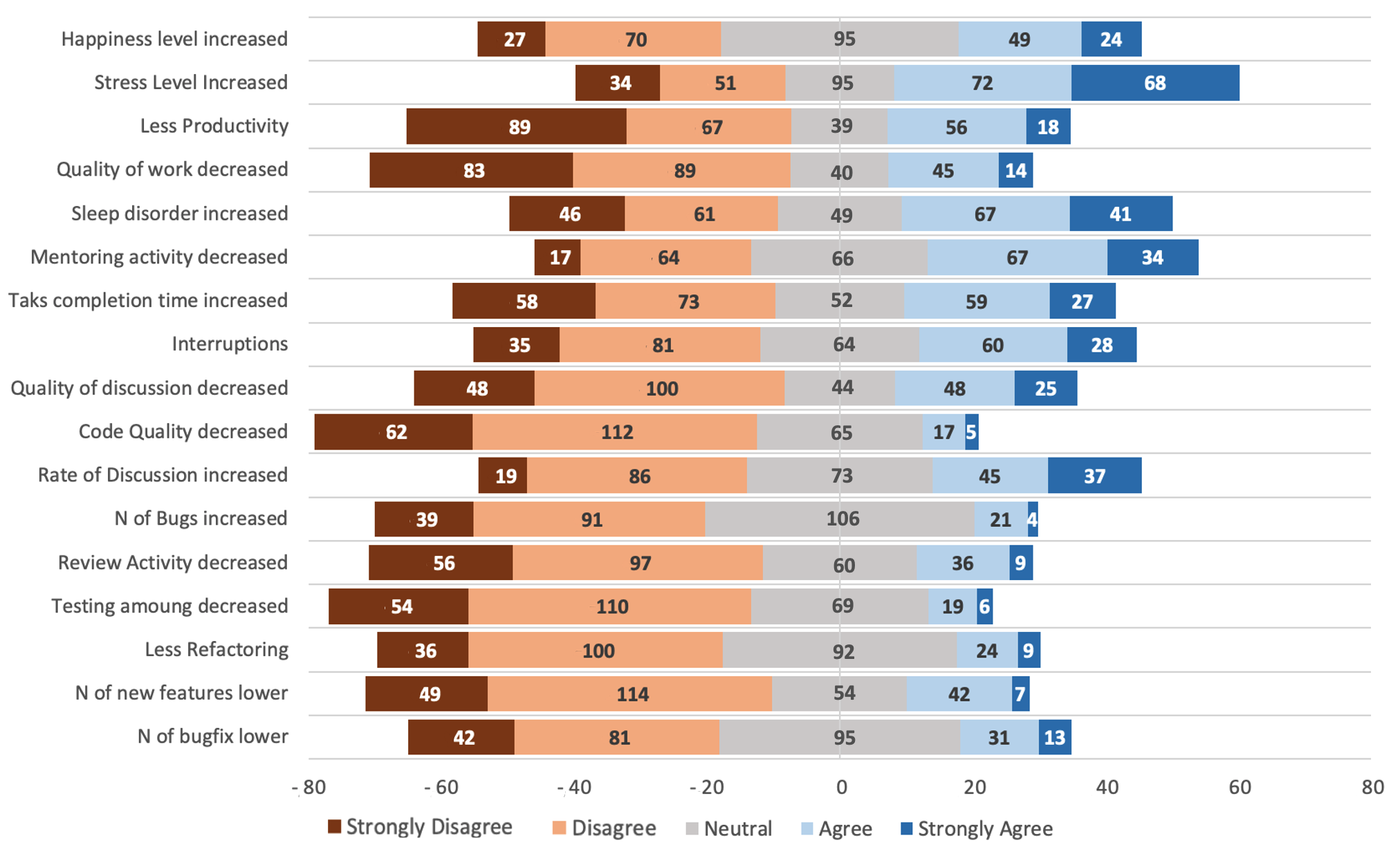}
  \caption{Survey Results Summary.}
  \label{fig:agree_disagree}
\end{figure}
\end{center}

We also investigated whether  interruptions are associated with number of children below 12 years of age living in the household. We analyzed the results from the survey and our results show that the average Likert score ranges from $2.0$ to $3.2$, indicating the perception that the number of children also does not have a major impact. On the contrary, people mostly disagree with this perception. Figure~\ref{fig:agree_disagree} shows the summary of the survey.

\observation{Majority of the respondents do not perceive the number of kids and people living in the house as a source of interruption.}

\section{Discussion}
\label{sec:discussionandimplication}

\vspace{0.2cm}\noindent{\bf Practitioners:} From our study, we found several suggestions for the practitioners to maintain a healthy work-life balance during a pandemic. Creating various ``forms of entertainment at home'' can help divert the mind away from all the negative news. Staying at home provides opportunities for practitioners to expand their knowledge and utilize this time to study and take online courses as several universities and platforms have their offerings. While pandemic has been tough for everyone, it has been positive for many with ``better sleep, more time with family, less travel time". Practitioners recommend taking ``more small breaks during the day to spend time with family and pets". Others mentioned the benefit of being disciplined in separating work time from rest/family time; for example, one respondent mentions: \textit{``No overtime. When my workday is done, I close my laptop and put in a drawer (``Out of sight, out of mind'') and get it out only on the next working day.''} Several respondents mention using the Pomodoro technique for time management, which uses a timer to break down work into intervals, has helped ``improve productivity''.

\vspace{0.2cm}\noindent{\bf Organizations:} Based on our analysis, we found that the COVID-19 situation does not necessarily result in reduced productivity and inferior code quality. Organizations can take active steps to help developers cope with COVID-19 and remain productive and produce high-quality code. We received many inputs from our respondents including the following: \textit{``On our company we have a permanent virtual room opened to attend doubts and have discussions.'', ``At work, my team has weekly 30min meetings to just hangout together on zoom. They were good at first to help us stay connected and sane'', ``Daily meetings during the beginning and end of the day, it allows visibility for the whole team and visibility about what is being executed. It was too efficient that the team is producing more'', ``My team opened up a remote call which everybody should be present during all day. This way we 'simulate' the real-world environment which we can talk with the other team members anytime. We noticed that the team collaboration has not decreased''.} From the above, we can note that organizations can positively impact their developers ability to cope well with COVID-19 in various ways to simulate their previous working environment (before the pandemic) and facilitate more interactions between the developers. In such a way, developers can effectively collaborate and feel less isolated in doing their tasks. 


\vspace{0.2cm}\noindent {\bf Researchers:} Our findings highlight that the impact of COVID-19 is not binary (reduce productivity vs. increase productivity) but rather a spectrum. For many of our observations, substantial proportions of respondents have differing opinions from each other. For example, for Observation 3 (\textit{Though the number of pull request comments created and updated decreased over the pandemic period, discussion among developers related to their work and review activity did not decrease during the pandemic.}), 37.63\% of respondents did not think there is an increment in discussion among team members since they began working from home in the pandemic period, with an average Likert score of 3.0 ("neutral"). 

More research is needed to uncover specific conditions that cause certain outcomes to be more prevalent. These conditions can correspond to personalities of different developers, their roles, their organization structure, and characteristics, their home conditions (e.g., ``low internet bandwidth'', ``irregular power supplies''), etc. To illustrate the impact of these factors, one of our respondents mentions ``... \textit{[I] have a low
need for social interaction. Sunbathing with the baby is enough to relieve the ``quarantine''} while another mentions ``\textit{decreased one-on-one conversations ... impacted our effectiveness''}. Yet another mention: ``\textit{Slightly increased the frequency of my panic attacks''}. This suggests that personality plays a role in the impact of COVID-19 to developers (c.f.,~\cite{XiaLBSL17} that also shows the impact of personality on the developer and team effectiveness prior to COVID-19). Developers with some physical/mental health conditions (e.g., panic attacks) may also be more adversely impacted by COVID-19.

The understanding gained from further research can result in: (1) creation of specific guidelines that can help developers or organizations adversely affected by COVID-19 to learn from other developers or organizations that have coped well with COVID-19, (2) Organizations to adopt different strategies to help developers of different personalities and conditions to cope with COVID-19. 

We do not view our study as a final definitive study, but rather one of the many that can shed ``full'' light into COVID-19 (or other pandemics), its impacts, and ways to mitigate those impacts. Some of the future studies can consider performing a smaller scale but more in-depth and focused study on a particular aspect (e.g., the impact of personality on how developers cope with COVID-19).

\section{Threats to Validity}
Our research findings may be subject to the concerns that we list below. We have taken all possible steps to reduce the impacts of these possible threats, but some could not be mitigated and it's possible that our mitigation strategies may not have been effective.

Our samples have been from a single source (Github) and single programming language (Java). This may be a source of bias, and our findings may be limited to open source programs from Github. However, we believe that the large number of projects sampled more than adequately addresses this concern. 


The set of analyzed metrics spans across 10 different categories and widely used in literature. However, we cannot guarantee that our set of metric are exhaustive. We plan to expand our metric set in future work. 

It is possible that there are defects in the implementation of our mining scripts. To that end, we have extensively tested our implementation, and manually verified sampled counts of different metrics.

Finally, it is possible that the survey participants misunderstood some of the survey questions. To mitigate this threat, we conducted a pilot study with developers with different experience levels from both open-source community and industry. We also conducted a pilot study with survey design experts. We updated the survey based on the findings of these pilot studies.

\section{Conclusions}

The COVID-19 pandemic has impacted the whole world in different ways. As software is still "eating the world", it is essential to understand COVID-19's impact on software projects. We conducted a mining software repository study based on 100 GitHub projects developed in Java using ten different metrics. Next, we surveyed 279 software development professionals from 32 countries for gathering more insights about the impact of COVID-19 on daily activities and wellbeing.

Based on our findings, we derived 12 observations that can be used by practitioners, organizations, and researchers. Practitioners can use our recommendations to maintain a healthy work-life balance during a pandemic. Organizations can learn from our survey respondents and take steps to remain productive while creating high-quality code. The research community can explore the social and human aspects to understand the impact of developer personality during a pandemic. 

As future work, we intend to expand the analysis of the projects (considering a wider time window) to better understand the impact of a year-long pandemic on software development.


\end{document}